\documentclass[conference]{IEEEtran}
\IEEEoverridecommandlockouts
\usepackage{cite}
\usepackage{amsmath,amssymb,amsfonts}
\usepackage{algorithmic}
\usepackage{graphicx}
\usepackage{textcomp}
\usepackage{xcolor}
\usepackage{multirow}
\usepackage{multicol}
\usepackage{makecell}

\usepackage{tikz,graphics,color,fullpage,float}

\usetikzlibrary{shadows,patterns,shapes,arrows,decorations.pathmorphing,backgrounds,positioning,fit,plotmarks,calc,spy,petri,topaths}
\usetikzlibrary{arrows.meta}
\usepackage{pgfplots}
\usepgfplotslibrary[groupplots]
\pgfplotsset{compat=1.18,
    /pgfplots/ybar legend/.style={
    /pgfplots/legend image code/.code={%
       \draw[##1,/tikz/.cd,yshift=-0.2em]
        (0cm,0cm) rectangle (4pt,0.6em);},
   },
}
\usetikzlibrary{matrix, patterns, patterns.meta, shapes.geometric}
\usepackage{pifont}
\usepackage{wrapfig}
\usepackage{pgfplotstable}
\usepackage[linesnumbered,titlenumbered,ruled,noend]{algorithm2e}

\newcommand*\smallcircled[1]{\tikz[baseline=(char.base)]{
            \node[shape=circle,draw,black,inner sep=0.5pt] (char) {#1};}}
\newcommand{\xmark}{\ding{55}}%
\usepackage{enumitem}
\usepackage{url}
\usepackage{balance}
\newcolumntype{C}[1]{>{\centering\arraybackslash}p{#1}}

\definecolor{blizzardblue}{rgb}{0.67, 0.9, 0.93} 
\definecolor{cadmiumgreen}{rgb}{0.0, 0.42, 0.24} 
\definecolor{lightgray}{rgb}{0.83, 0.83, 0.83} 
\definecolor{deepskyblue}{rgb}{0.0, 0.75, 1.0} 
\definecolor{forestgreen}{rgb}{0.13, 0.55, 0.13} 
\definecolor{lightcyan}{rgb}{0.88, 1.0, 1.0} 
\definecolor{limegreen}{rgb}{0.2, 0.8, 0.2} 
\definecolor{mediumblue}{rgb}{0.0, 0.0, 0.8} 
\definecolor{seagreen}{rgb}{0.18, 0.55, 0.34} 
\definecolor{darkgoldenrod}{rgb}{0.72, 0.53, 0.04}
\definecolor{darkmagenta}{rgb}{0.55, 0.0, 0.55}
\definecolor{darkseagreen}{rgb}{0.56, 0.74, 0.56}
\definecolor{cerulean}{rgb}{0.0, 0.48, 0.65}
\definecolor{darkseagreen}{rgb}{0.56, 0.74, 0.56}
\definecolor{darkspringgreen}{rgb}{0.09, 0.45, 0.27}
\definecolor{darkturquoise}{rgb}{0.0, 0.81, 0.82}
\definecolor{denim}{rgb}{0.08, 0.38, 0.74}
\definecolor{dollarbill}{rgb}{0.52, 0.73, 0.4}
\definecolor{darkgreen}{rgb}{0.0, 0.2, 0.13}
\definecolor{darkblue}{rgb}{0.0, 0.0, 0.55}
\definecolor{calpolypomonagreen}{rgb}{0.12, 0.3, 0.17}
\definecolor{darkcyan}{rgb}{0.0, 0.55, 0.55}
\definecolor{ferngreen}{rgb}{0.31, 0.47, 0.26}
\definecolor{forestgreen(traditional)}{rgb}{0.0, 0.27, 0.13}
\definecolor{navyblue}{rgb}{0.0, 0.0, 0.5} 
\definecolor{forestgreen}{rgb}{0.0, 0.27, 0.13} 
\definecolor{burgundy}{rgb}{0.5, 0.0, 0.13} 
\definecolor{slategray}{rgb}{0.44, 0.5, 0.56} 
\definecolor{charcoal}{rgb}{0.21, 0.27, 0.31} 
\definecolor{maroon}{rgb}{0.5, 0.0, 0.0} 
\definecolor{darkteal}{rgb}{0.0, 0.2, 0.2} 
\definecolor{taupe}{rgb}{0.28, 0.24, 0.2} 
\definecolor{olivegreen}{rgb}{0.33, 0.42, 0.18} 
\definecolor{steelblue}{rgb}{0.27, 0.51, 0.71} 
\definecolor{bluebell}{rgb}{0.301, 0.341, 0.690} 
\definecolor{aliceblue}{rgb}{0.953, 0.953, 0.953} 
\definecolor{vermilion}{rgb}{0.988, 0.392, 0.298} 
\definecolor{corduroy}{rgb}{0.600, 0.600, 0.600} 
\definecolor{brickred}{rgb}{0.804, 0.416, 0.518} 
\definecolor{eucalyptus}{RGB}{59, 110, 100}
\definecolor{cadet}{RGB}{66, 123, 114}
\definecolor{spray}{RGB}{141, 192, 179}
\definecolor{portage}{RGB}{123, 129, 164}
\definecolor{palelilac}{RGB}{209, 200, 218}
\definecolor{laurelgreen}{RGB}{137, 138, 124}
\definecolor{satPointsColor}{HTML}{D7191C}
\definecolor{unsatPointsColor}{HTML}{FDAE61}
\definecolor{timedOutPointsColor}{HTML}{ABDDA4}
\definecolor{rendering}{HTML}{2B83BA}
\definecolor{lightBlue}{RGB}{173, 216, 230}   
\definecolor{lightRed}{RGB}{255, 182, 193}    
\definecolor{lightYellow}{RGB}{255, 255, 224} 
\definecolor{lightGreen}{RGB}{144, 238, 144}  
\definecolor{purple}{RGB}{128, 0, 128}        
\definecolor{gray}{RGB}{128, 128, 128}        
\definecolor{deeperBlue}{RGB}{100, 149, 237}  
\definecolor{deeperRed}{RGB}{220, 20, 60}     
\definecolor{deeperYellow}{RGB}{238, 220, 130} 
\definecolor{deeperGreen}{RGB}{34, 139, 34}   
\definecolor{deeperPurple}{RGB}{104, 34, 139} 
\definecolor{deeperGray}{RGB}{105, 105, 105}  
\definecolor{richYellow}{RGB}{204, 204, 0}  
\definecolor{deeperorange}{RGB}{205,85,0}
\definecolor{deeppeach}{rgb}{1.0, 0.8, 0.64}
\definecolor{viridian}{rgb}{0.25, 0.51, 0.43}
\definecolor{manatee}{rgb}{0.59, 0.6, 0.67}
\definecolor{darkbrown}{rgb}{0.4, 0.26, 0.13}
\pdfoutput=1
\usepackage[table]{xcolor}

\usepackage[a4paper, total={184mm,239mm}]{geometry}
\def\BibTeX{{\rm B\kern-.05em{\sc i\kern-.025em b}\kern-.08em
    T\kern-.1667em\lower.7ex\hbox{E}\kern-.125emX}}
\begin{document}

\title{DAOP: \underline{D}ata-\underline{A}ware \underline{O}ffloading and Predictive \underline{P}re-Calculation for Efficient MoE Inference
\thanks{\textit{Proceedings of the DATE Conference}, Lyon, France, 2025. Copyright 2025 by the author(s).}
}

\author{
    \IEEEauthorblockN{Yujie Zhang, Shivam Aggarwal, Tulika Mitra}
    \IEEEauthorblockA{\textit{School of Computing, National University of Singapore}}
    \IEEEauthorblockA{\{zyujie, shivam, tulika\}@comp.nus.edu.sg}
}
\maketitle

\begin{abstract}
Mixture-of-Experts (MoE) models, though highly effective for various machine learning tasks, face significant deployment challenges on memory-constrained devices. While GPUs offer fast inference, their limited memory compared to CPUs means not all experts can be stored on the GPU simultaneously, necessitating frequent, costly data transfers from CPU memory, often negating GPU speed advantages. To address this, we present DAOP, an on-device MoE inference engine to optimize parallel GPU-CPU execution. DAOP dynamically allocates experts between CPU and GPU based on per-sequence activation patterns, and selectively pre-calculates predicted experts on CPUs to minimize transfer latency. This approach enables efficient resource utilization across various expert cache ratios while maintaining model accuracy through a novel graceful degradation mechanism. Comprehensive evaluations across various datasets show that DAOP outperforms traditional expert caching and prefetching methods by up to 8.20$\times$ and offloading techniques by 1.35$\times$ while maintaining accuracy. 
\end{abstract}
\begin{IEEEkeywords}
MoE inference engine, data-aware, expert offloading, GPU-CPU hybrid execution
\end{IEEEkeywords}

\section{Introduction}
\begin{figure}[t]
    \centering
   \begin{minipage}{0.41\columnwidth}
     \centering
     \includegraphics[width=\columnwidth]{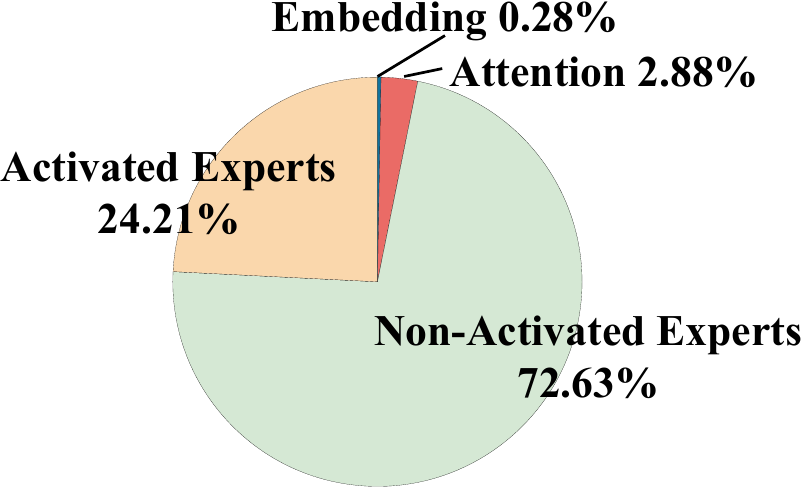}
     \vspace*{-0.6cm}
     \caption{Average parameter distribution in Mixtral 8x7B~\cite{jiang2024mixtral}.}\label{fig:moe_dist}
   \end{minipage} \hspace*{0.3cm}
   \begin{minipage}{0.41\columnwidth}
     \centering
     \includegraphics[width=\columnwidth]{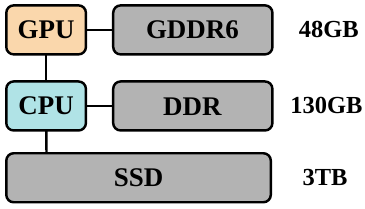}
     \caption{NVIDIA RTX A6000 GPU specifications.}\label{fig:a6000_gpu}
   \end{minipage} 
\end{figure}

Mixture-of-Experts (MoE) architecture~\cite{jacobs1991adaptive} addresses the substantial computational demands of Large Language Models (LLMs) by utilizing multiple expert networks and activating only a subset for each input. 
For instance, in the widely-used Mixtral 8x7B MoE model~\cite{jiang2024mixtral} with 46.6 billion parameters, only 27.4\% of the parameters are activated per input sequence (see Fig.~\ref{fig:moe_dist}). 
This sparse activation balances model capacity with computational efficiency, significantly reducing computational load compared to traditional dense models.

However, deploying these models in low-resource environments is challenging due to limited GPU memory capacity (Fig.~\ref{fig:a6000_gpu}), which makes it difficult to store all experts in GPU memory and leverage fast execution. Consequently, experts must frequently be loaded from CPU memory, introducing substantial overhead that negates the speed advantages of GPU execution. This data transfer severely undermines inference speed, a critical factor for real-time responsiveness.

Quantization and sparsity techniques~\cite{aggarwal2024shedding,lu2024not,10546782} can reduce computational and memory demands but often result in inconsistent performance across various machine learning tasks. Recent strategies like caching frequently activated experts on GPUs and prefetching aim to mitigate lengthy data transfers between CPUs and GPUs~\cite{du2024sida,hwang2024pre,zhong2024adapmoe}. However, significant delays in computation and expert loading make it challenging to mask the transfer bottlenecks, especially in large MoE models. Innovative strategies are required to utilize CPU and GPU resources to accelerate inference effectively.

As detailed in Table~\ref{tab:block_expload_compare}, on the NVIDIA A100 platform, migrating a single expert within a transformer block from the CPU to the GPU is approximately 32$\times$ slower than executing the entire block on the GPU with all parameters loaded.
This disparity fundamentally limits the efficacy of expert prefetching in concealing memory transfer overhead within the compute-I/O pipeline: prefetching is only effective when host-device transfer time is subsumed by the available computation window, yet large expert weight transfers routinely exceed this window under memory-constrained MoE inference.
Conversely, the activations required to be transferred between CPU and GPU for an expert execution on CPU are approximately one ten-thousandth the size of expert weights. Therefore, for an expert that is not present in GPU memory, offloading expert execution to CPU resources is more feasible than loading experts to the GPU for computation.

\setlength\tabcolsep{2.5pt} 
\begin{table}[t]
    \centering
    \caption{
    Execution times (ms)$^{\mathrm{a}}$ for Transformer block operations$^{\mathrm{b}}$ and expert migration in Mixtral 8x7B, tested on an NVIDIA A100 GPU and an Intel Xeon Gold 6326 CPU @ 2.9GHz, connected via a PCIe 4.0 interface with 64 GB/s bandwidth
    }
    \label{tab:block_expload_compare}
        \begin{tabular}{p{0.9cm}|p{0.9cm}|c|c} 
        \hline            
            \multicolumn{2}{c|}{\textbf{Block Execution}} & \multirow{2}{*}{\textbf{\makecell{Expert Migration\\(CPU $\rightarrow$ GPU)}}} & \multirow{2}{*}{\textbf{\makecell{Expert Input / Output\\Transition (CPU $\leftrightarrow$ GPU)}}} \\
            \cline{1-2}
            \centering\textbf{CPU} & \centering\textbf{GPU} & & \\
            \hline
            \centering 8.02 & \centering 1.24 & \textcolor{red}{\textbf{39.87}} & 0.02 \\
        \hline
        \end{tabular}
        \\
        {\raggedleft\footnotesize {{$^{\mathrm{a}}$The time is measured during the decode stage with input/output length 256.} \\
        $^{\mathrm{b}}$Each block includes a self-attention mechanism and expert (linear) layers.}}
\end{table}

We introduce \emph{DAOP}, a data-aware MoE inference engine optimized for low-resource devices.
DAOP dynamically offloads non-critical experts to CPUs based on inference task requirements, predicting expert activations one layer in advance and selectively pre-calculating them on CPUs using approximate methods. 
This approach enables DAOP to leverage parallel computing across CPUs and GPUs, significantly accelerating inference with minimal impact on accuracy. 
Notably, it eliminates the need for model modification or fine-tuning, providing a distinct advantage over task-specific quantization techniques.

Our key contributions are summarised as follows: 
\begin{itemize}[noitemsep,topsep=0pt]
    \item We present and analyze various design choices crucial for large-scale MoE model inference on memory-constrained GPU devices.
    \item We develop a data-aware MoE inference engine, DAOP, to adaptively and efficiently utilize CPU and GPU resources, thus alleviating GPU memory constraints and I/O bottlenecks. 
    \item We thoroughly evaluate DAOP using popular Mixtral and Phi MoE models across various datasets on NVIDIA RTX A6000. Despite varying numbers of experts cached on the GPU, our algorithm sustains acceptable accuracy and consistently surpasses expert caching and prefetching approaches by 8.20$\times$ and offloading methods by 1.35$\times$.
\end{itemize}

\section{Preliminaries}

\subsection{LLMs with Mixture-of-Experts (MoE)}
\begin{wrapfigure}{li}{0.25\textwidth}
  \centering
    \includegraphics[width=0.25\textwidth]{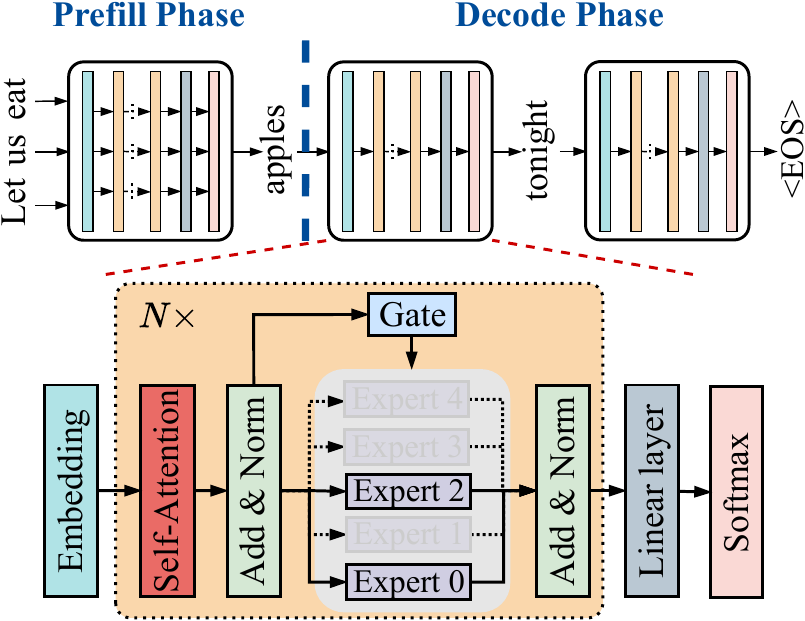}
    \caption{The decoder-only MoE-based LLM inference procedure with top-2 experts activated per token.}
    \label{fig:prefill-decode}
\end{wrapfigure}
This study examines decoder-only MoE-based LLMs. Fig.~\ref{fig:prefill-decode} illustrates the inference process of an example model activating top-2 experts per token.

\noindent \textbf{Decoder-only Architecture.} Prominently used in leading LLMs such as OpenAI's GPT series~\cite{kalyan2023survey} and Meta's LLaMA~\cite{touvron2023llama}, this architecture processes input sequences in two phases: \textit{prefill} and \textit{decode}. In the prefill stage, tokens embedded with positional information traverse the model to generate the initial output token. The subsequent decode phase employs an autoregressive mechanism to generate each following token sequentially, extending processing time proportionally to the output sequence length. The model comprises multiple transformer blocks, each featuring self-attention, expert layers (feed-forward networks), normalization, and residual connections. Self-attention elucidates inter-word relationships, and expert layers perform non-linear transformations to uncover complex data patterns.

\noindent \textbf{MoE Structure.} As illustrated in Fig.~\ref{fig:prefill-decode} (bottom), MoEs activate only a fixed select group of experts (e.g., top-1 or top-2) for each token. 
The MoE mechanism encompasses \textit{expert selection} and \textit{execution}. A \textit{gating function}, implemented as a multi-layer perceptron (MLP), assesses the hidden states following self-attention to determine each expert's activation probabilities, indicating their importance to the specific token. Once selected, these experts process and transform the input tokens into output tokens.
For clarity, this paper refers to computations within the block, excluding expert selection and computation, as the non-MoE part. 

\subsection{Related Work}
\label{sec:related_work}
Edge inference techniques for MoE models often employ expert caching and prefetching strategies.
For instance, Mixtral-Offloading~\cite{eliseev2023fast} dynamically caches expert resources by offloading least used experts to host memory and uploading activated experts to GPU devices, enhanced by mixed quantization to accelerate inference.
MoE-Infinity~\cite{xue2024moe} employs activation-aware prefetching to mitigate I/O delays based on sequence-level expert activation patterns.
EdgeMoE~\cite{yi2023edgemoe} reduces expert size through expert-specific bit-width adaptation with minimal accuracy loss and preloads anticipated experts using a compute-I/O pipeline.
SiDA-MoE~\cite{du2024sida} uses a hash function to predict expert activation patterns, allowing efficient preloading during runtime.
Pre-gated MoE~\cite{hwang2024pre} introduces a predictive pre-gating mechanism for optimized prefetching.
Despite these advancements, caching and prefetching strategies often struggle to mask expert loading overhead effectively in MoE models with large-scale experts, where limited memory bandwidth remains a bottleneck. Other works, such as Fiddler~\cite{kamahori2024fiddler}, minimize data movement by executing experts directly on the CPU. Nevertheless, its approach is simplistic and lacks adaptive exploitation of expert activation patterns to enhance model execution efficiency further.
\section{Observations \& Insights}
\label{sec:findings}
This section presents various observations and design choices w.r.t MoE model inference on memory-constrained GPU-CPU devices. 
\begin{wrapfigure}{li}{0.2\textwidth}
  \raggedright
    \includegraphics[width=0.2\textwidth]{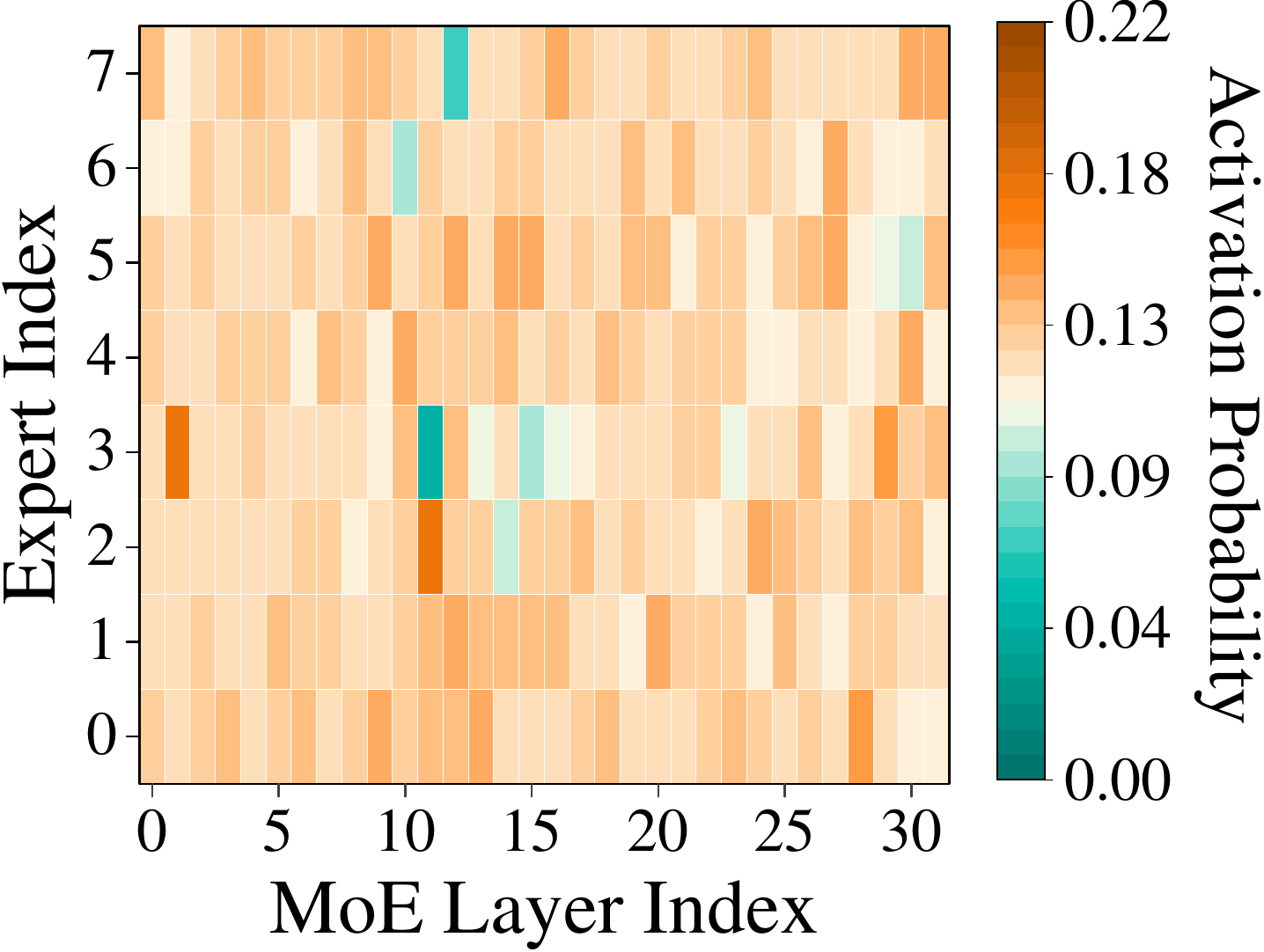}
    \caption{Layer-wise expert activation pattern of Mixtral 8x7B on dataset C4.}
    \label{fig:expt-pred-c4}
\end{wrapfigure}
\noindent\smallcircled{1} \textbf{Dominant experts inherently vary with input sequences.}
In MoE structures, experts are trained to achieve balanced activation frequencies across various tasks~\cite{fedus2022review}. 
This balance leads to inherent variability in dominant experts for different inputs, resulting in cache misses and frequent migrations of experts between memory hierarchies.
For example, in a comprehensive dataset such as C4, experts are activated with nearly uniform probability across the dataset~\cite{lu2024not}, as illustrated in Fig.~\ref{fig:expt-pred-c4}. This leads to balanced activation within individual input sequences. Alternatively, across different sequences, significant deviations in expert preference frequently occur. Consequently, limited GPU memory struggles to capture activation locality to reduce frequent expert migration. Therefore, adaptive and dynamic offloading mechanisms are essential for efficient execution.

\noindent\smallcircled{2} \textbf{Expert activation patterns exhibit substantial similarity between prefill and decode phases across input sequences.}  
To quantify this observation, we use an $L \times E$ matrix to represent these patterns, where $L$ represents the number of MoE layers and $E$ denotes the number of experts per layer. 
Within this matrix, $P_{i,j}$ and $D_{i,j}$ signify the activation probabilities of expert $j$ at layer $i$ during the prefill and decode stages, respectively. These probabilities are determined by the ratio of tokens routed to expert $j$ at layer $i$ to the total tokens processed by that layer. The similarity between the expert activation matrices for two phases is quantified by averaging the cosine similarities of corresponding rows in the two matrices, as detailed as follows: 
\begin{equation}
\label{eq:dist}
\hspace{0pt}\text{Similarity} \left(P,D\right) = \frac{1}{L} \sum_{l=1}^{L} \text{CosineSimilarity} \left(P_{l}, D_{l}\right)
\end{equation}

Table~\ref{tab:pre_dec_dist} illustrates the average similarity for the Mixtral 8x7B model, calculated individually for three datasets: C4, MATH, and GSM8K. We randomly select 512 samples for each dataset, with the overall similarity across all datasets averaging approximately 90.72\%. 
Consequently, in our design, the expert activation pattern during the prefill stage informs the optimal distribution of experts between CPU and GPU resources, allowing less-utilized experts to be offloaded to the CPU for approximate computing, thereby minimizing impacts on model accuracy.
\setlength\tabcolsep{2pt} 
\begin{table}[t]
    \centering
    \caption{Average similarity of expert activation matrices between prefill and decode phases in the Mixtral 8x7B model}
    \label{tab:pre_dec_dist}
        \begin{tabular}{c|c|c|c|c}
        \hline
            & \textbf{C4} & \textbf{MATH} & \textbf{GSM8K} & \textbf{Average}\\
            \hline
            \textbf{Similarity (\%)} & 90.05 & 90.37 & 91.74 & 90.72\\
        \hline
        \end{tabular}
\end{table}
\begin{figure}[t]
    \raggedleft

\begin{tikzpicture}
    \centering   
    \begin{groupplot}[
        group style={
        group name=my plots,
        group size=1 by 1,
        xlabels at=edge bottom,
        xticklabels at=edge bottom,
        vertical sep=0pt,
        horizontal sep=0pt,},
        xtick=data,
        x tick label style={font=\tiny,rotate=0,text height=2pt,text width=1.5cm,align=center,inner sep=2pt,outer sep=2pt},
        xticklabels from table={data/expt_pred.dat}{Layer_Index},
        xlabel style={font=\scriptsize, align=center, text height=0pt,inner sep=1pt,outer sep=1pt},
        ytick pos=left,
        y tick label style={font=\tiny,rotate=0,text height=0pt,inner sep=1.5pt,outer sep=0pt},
        ylabel style={font=\scriptsize, align=center, text height=0.1pt,inner sep=0pt,outer sep=0pt},]
        
        \nextgroupplot[width=\columnwidth,, height=3.3cm, 
                        ylabel={Prediction\\Accuracy (\%)},
                        xlabel={Layer Index},
                        ymin=55,
                        ymajorgrids=true,
                        grid style=densely dotted,
                        enlarge x limits={abs=0.5},
                        ytick={0,10,20,30,40,50,60,70,80,90,100},
                        yticklabels={0,10,20,30,40,50,60,70,80,90,100},
                        xtick={0,1,2,3,4,5,6,7,8,9,10,11,12,13,14,15,16,17,18,19,20,21,22,23,24,25,26,27,28,29,30,31},
                        xticklabels={1,,,,5,,,,,10,,,,,15,,,,,20,,,,,25,,,,,30,},
                        extra y ticks={80},
                        extra tick style={grid=major,major grid style={red, densely dashed}},
                        extra y tick labels={},
                        typeset ticklabels with strut,
                        legend columns=4,
                        legend style={at={(0.45,1.05,1)}, anchor=south, font=\scriptsize, draw=none,
                        legend image post style={mark size=1.9pt}},grid]

        \addplot+[mark=o, deeperBlue, mark size=1.4pt,, mark options={line width=0.6pt, solid, fill=deeperBlue},]table[x expr = \coordindex,y = Alpaca,]{data/expt_pred.dat};
        \addlegendentry{Alpaca}

        \addplot+[mark=o, darkgoldenrod, mark size=1.4pt, mark options={line width=0.6pt, solid, fill=darkgoldenrod},]table[x expr = \coordindex,y = Math,]{data/expt_pred.dat};
        \addlegendentry{MATH}
        


        \addplot+[mark=o, deeperGreen, mark size=1.4pt, mark options={line width=0.6pt, solid, fill=deeperGreen},]table[x expr = \coordindex,y = C4,]{data/expt_pred.dat};
        \addlegendentry{C4}
        
        \draw (8,65) node[align=center] {\scriptsize \bfseries \textcolor{red}{Avg. Accuracy: 84.11\%}};
        
    \end{groupplot}           
\end{tikzpicture}
    \caption{Layer-wise expert prediction accuracy for the Mixtral 8x7B model, one layer ahead, during the decode phase.}
    \label{fig:expt-pred-acc}
\end{figure}
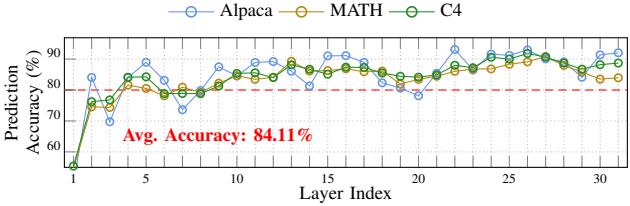

\noindent\smallcircled{3} \textbf{Expert prediction accuracy is generally high one layer in advance.} 
To forecast the experts needed for the next layer, we apply the subsequent layer’s gating function to the same hidden states employed by the current MoE layer's gating function. Transformer layers, designed to be residual, naturally enhance each successive layer’s hidden states, providing a solid foundation for predicting future states.
Our empirical results, depicted in Fig.~\ref{fig:expt-pred-acc}, reveal that the layer-wise expert prediction accuracy for the Mixtral 8x7B model, when averaged across the Alpaca, MATH, and C4 datasets, stands at 84.11\%. These datasets, which span mathematics and world knowledge, underscore the broad applicability of this prediction mechanism.
Thus, high prediction accuracy ensures the effectiveness of expert pre-calculation in reducing GPU waiting times.

\section{DAOP: MoE Inference Engine}
\begin{figure}[t]
    \centering
    \includegraphics[width=0.46\textwidth]{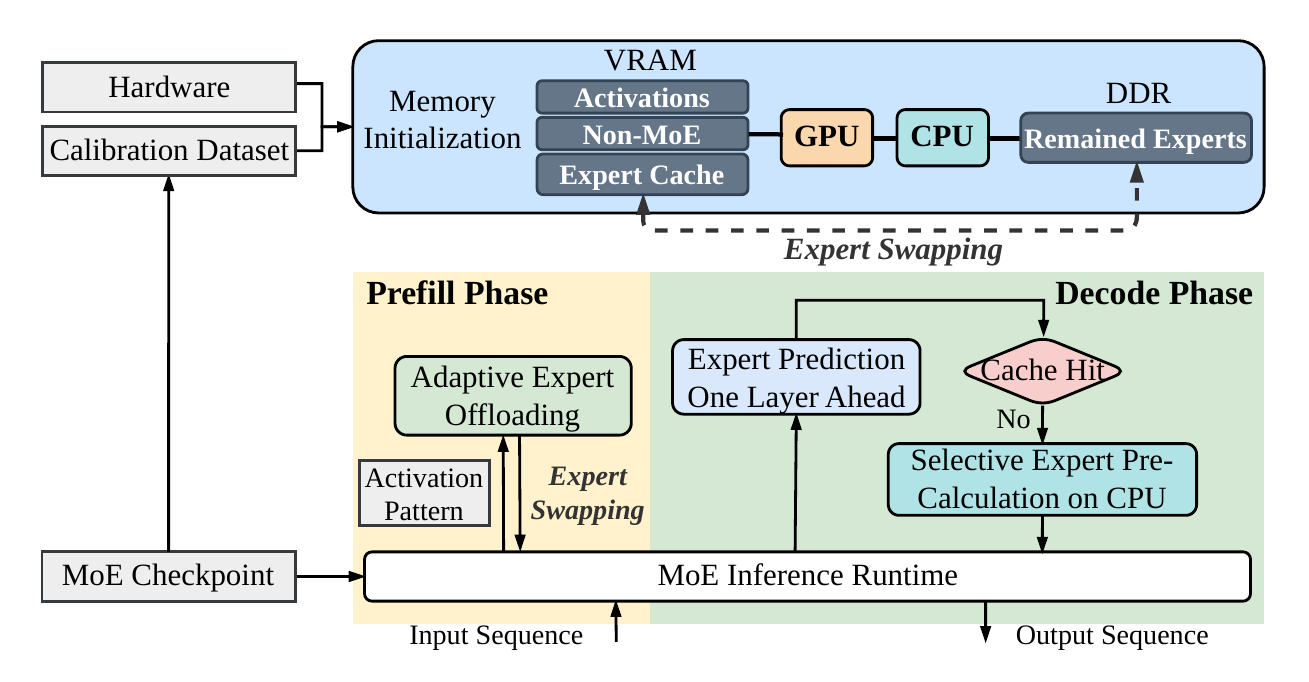}
    \caption{Design Overview of DAOP.}
    \label{fig:algo_ours}
\end{figure}

We now outline \emph{DAOP} based on the insights 
from the observations earlier. 
Fig.~\ref{fig:algo_ours} illustrates the critical components: (a) sequence-specific expert allocation between CPUs and GPUs (based on observations \smallcircled{1} \& \smallcircled{2}) and (b) a selective expert pre-calculation strategy driven by predictions without necessitating model fine-tuning (based on observation \smallcircled{3}).
We dynamically offload non-dominant experts to CPU memory for each sequence and effectively conceal expert migration overhead during the prefill phase. 
In the decode phase, we utilize expert parallelism based on prediction to leverage CPU resources, accelerating inference speed.

\subsection{Memory Initialization} 
We initially allocate non-MoE and dominant experts to the GPU to maximize memory utilization, with others assigned to the CPU. 
Utilizing the calibration dataset, we identify the expert activation pattern through the decode phase and prioritize experts based on layer-wise activation probabilities for GPU assignment. 
We standardize the expert cache size across all layers, ensuring it contains experts with the highest activation probabilities. Subsequently, any remaining space—smaller than the layer count—accommodates additional experts, prioritized by their activation frequency.

\subsection{Sequence-specific Expert Allocation} 
We determine the expert activation pattern for each input sequence during the prefill stage and dynamically adjust expert allocation to optimize reuse during the decode stage. To minimize I/O overhead, expert migration is restricted to the prefill stage for cache adjustments, with this configuration maintained throughout decoding.

Algorithm~\ref{alg:exp_alloc} details our sequence-specific expert allocation method. After the gating function to determine the expert activation pattern in each block \textit{(Line 4)}, we count the tokens assigned to each expert, treating this as the expert's activity level \textit{(Line 7-8)}.
We generate tuples, grouping the most active CPU experts with the least active GPU experts \textit{(Line 5-9)}. 
If a GPU expert has fewer assigned tokens than its CPU counterpart, we swap their locations \textit{(Line 11-13)}. 
We define a comparison threshold, $SwapInOut$, as 1.05 to avoid unnecessary swaps when the token counts are similar.
This process is repeated for all tuples. Finally, we offload less-utilized experts to the CPU and move the most-utilized ones to the GPU. 
Next, we input multiple tokens from the input sequence to the experts, whether on the CPU or GPU, for parallel local execution \textit{(Line 14)}. While assigning multiple tokens to each expert increases GPU utilization and processing speed, it also proportionally raises CPU execution time. Therefore, placing highly utilized experts on GPUs improves inference efficiency and helps mask expert allocation overhead.

\begin{algorithm}[t]
\small
    \caption{Sequence-specific Expert Allocation}
    \label{alg:exp_alloc}
\KwData{initialized expert cache, model blocks, 
comparison threshold SwapInOut}
\KwResult{updated expert cache, firstly generated output token}
    inputTokens = Embedding(inputTokens);\\
    \ForEach {block in blocks}{
        inputTokens = executeNonMoE(inputTokens);\\
        \{RouteInfo\} = gating(inputTokens);\label{algline:gating}\\
        SwapNum = 0.5 $*$ getNum(block.Experts);\\
        \{ExpsGPU, ExpsCPU\} = findExpertLocs(layer.Experts);\\
        HotExps = getTopKActiveExperts(ExpsCPU, SwapNum);\label{algline:token_activity_first}\\
        ColdExps = getBottomKActiveExperts(ExpsGPU, SwapNum);\label{algline:token_activity_second}\\
        \ForEach {HotExp, ColdExp in zip(HotExps, ColdExps)}{ 
            \{HotProb, ColdProb\} = getActiveProb(HotExp, ColdExp);\\
            \If{HotProb $\geq$ SwapInOut $*$ ColdProb}{
                swap an expert ColdExp out the expert cache;\\
                swap an expert HotExp in the expert cache;\\
            }\label{algline:swapping}
        }\label{algline:traverse_tuples}
        local expert execution for tokens in inputTokens in parallel;\\
    }
    outputToken = Normalization(inputTokens)
     

\end{algorithm}

\subsection{Prediction-based Expert Pre-Calculation} Fig.~\ref{fig:pre_cal} illustrates an example computation flow of expert pre-calculation based on prediction during the decode phase. 
In this stage, the selected experts are executed in parallel in their current locations to minimize time-consuming I/O overhead and accelerate inference, eliminating the need for reallocation.  
Given the execution time difference between CPUs and GPUs, we predict the experts required for the next layer. When the predicted experts are on the CPU, we selectively pre-calculate their execution using the hidden states from the non-MoE computation in the current layer. If the predicted experts are on the GPU, we feed the hidden states directly from the same layer’s non-MoE execution to the experts. This approach allows CPU-based experts to begin execution immediately after the non-MoE calculation in the previous block, providing sufficient time to parallelize expert execution across both CPU and GPU. Key features of this strategy include:
\begin{figure}[t]
    \centering
    \includegraphics[width=0.48\textwidth]{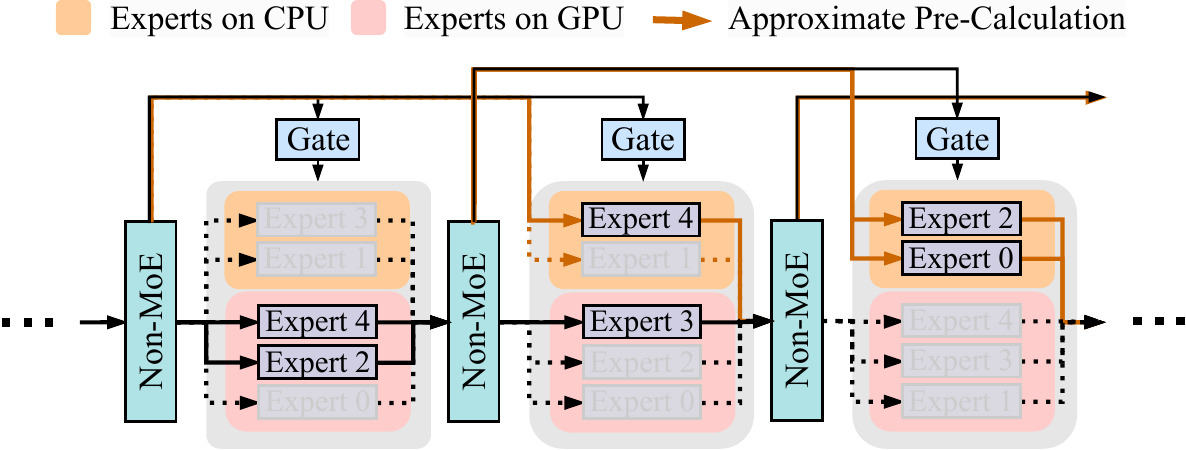}
    \caption{Computation flow of expert pre-calculation based on prediction.}
    \label{fig:pre_cal}
\end{figure}

\paragraph{Accuracy Stabilization} As shown in Fig.~\ref{fig:expt-pred-acc}, expert prediction accuracy stabilizes and becomes satisfactory after the first few layers.
Consequently, we use the gate function of the $(i+1)$-th block on the activations from the non-MoE computation of the $i$-th block to predict expert selection for block $(i+1)$, applicable when $ i\ge 4$. 
For blocks where $0\le i<4$, we use the original gate function to determine the activated experts. When the activated experts are already offloaded to CPUs, we transmit the hidden states generated by the non-MoE part to CPUs and send expert computation results back to GPUs, as the experts' input and output sizes are significantly smaller than the expert parameters.

\paragraph{Graceful Degradation}  
If two predicted experts are on CPUs, we replace the one with the lower score with the next-best expert already on the GPU to minimize overhead. Even if the next-best expert exhibits a lower score, its input data is generated from the non-MoE computation of the current block, leading to a high contribution to outputs. If no suitable alternative is available, the original selection is maintained for execution, as illustrated in the third block of Fig.~\ref{fig:pre_cal}.

\begin{figure}[t]
    \centering
    \includegraphics[width=0.48\textwidth]{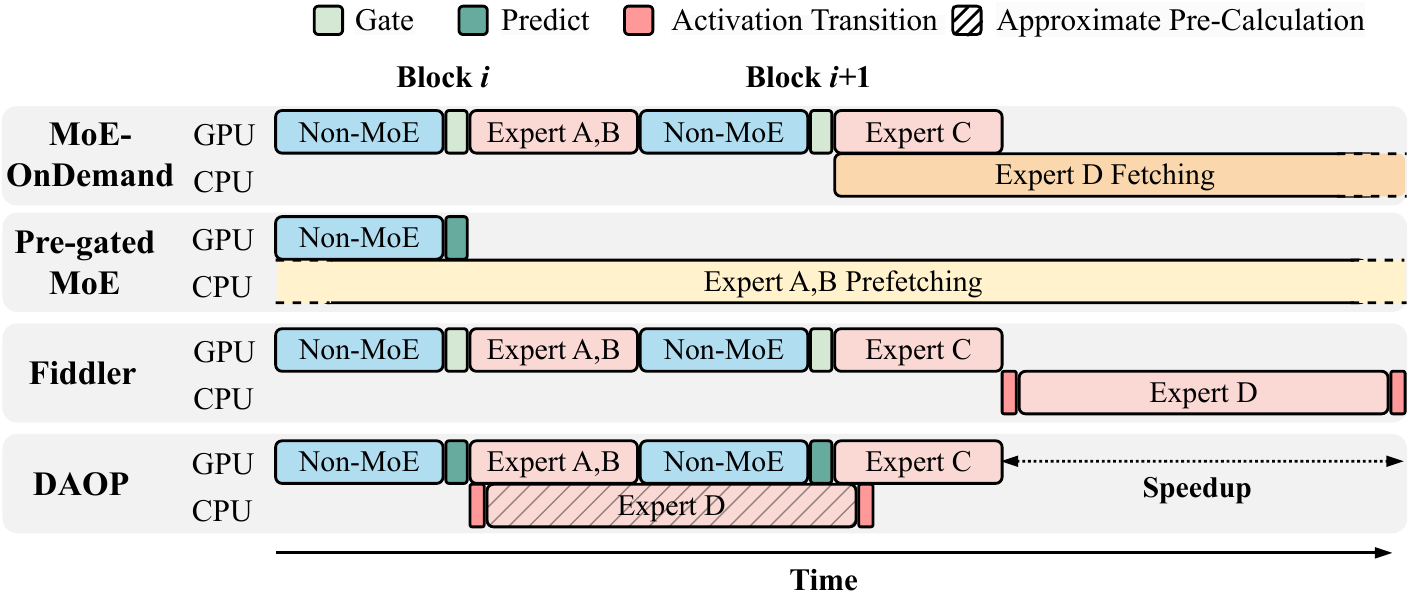}
    \caption{Execution timeline of DAOP and related works, showing two consecutive Transformer blocks with experts A and B activated for the first block and experts C and D for the second. Initially, experts A, B, and C are cached on GPUs for MoE-OnDemand, Fiddler, and DAOP due to high activation probabilities calibrated.}
    \label{fig:execution_timeline}
\end{figure}

\paragraph{GPU-CPU Hybrid Execution} Fig.~\ref{fig:execution_timeline} shows the execution timeline of DAOP and related methods during the decode stage. 
MoE-OnDemand places non-MoE and dominant experts on the GPU, with the rest stored in host memory. During runtime, experts not on the GPU are migrated dynamically, incurring significant overhead.
In both MoE-OnDemand and Pre-gated MoE, fetching or prefetching experts causes severe I/O bottlenecks, as large expert sizes make it difficult to mask I/O latency during block execution.
Conversely, Fiddler and DAOP offload expert execution to the CPU when needed experts are absent from the GPU, avoiding migration overhead. 
DAOP further improves upon Fiddler by leveraging parallel execution between the CPU and GPU, pre-calculating experts on the CPU one layer in advance, thereby accelerating inference. This approach relies on accurate expert prediction, ensuring minimal impact on overall model performance.
\section{Experimental Evaluation}
\subsection{Experimental Setup}
\paragraph{Models and Datasets} 
We assess the performance of DAOP using two popular decoder-only sparse MoE models, both employing the representative top-2 activation mechanism, as detailed in Table~\ref{tab:moe_models}.
This mechanism is extensively employed and considered representative in recently proposed MoE models, as highlighted in a comprehensive survey~\cite{Cai_2025}.
The evaluation datasets span several domains, including commonsense reasoning datasets like HellaS (10-shot)~\cite{zellers2019hellaswag}, Arc-e, Arc-c (25-shot)~\cite{clark2018think}, PIQA~\cite{bisk2020piqa}, WinoG (5-shot)~\cite{sakaguchi2021winogrande}, and TruthfulQA~\cite{lin2022truthfulqa}, all in a 0-shot format except where noted. Additionally, we assess popular aggregated results datasets such as MMLU (5-shot)~\cite{hendrycks2020measuring} and BBH (3-shot)~\cite{suzgun2023challenging}, alongside world knowledge and math datasets including TriviaQA (0-shot)~\cite{joshi2017triviaqa} and GSM8K (5-shot)~\cite{cobbe2021training}. 
The evaluation metrics for summarization are Rouge-1 and Rouge-2 scores~\cite{lin2004rouge}, and for question answering, ExactMatch scores. 

\paragraph{Hardware} 
We evaluate the performance of DAOP on an edge platform configured with one Intel i9-10980XE CPU and an NVIDIA RTX A6000 GPU. The CPU includes 18 physical cores at 3.0GHz and 130GB of host memory, while the GPU comprises 48GB GDDR6 with a memory bandwidth of 768GB/s, utilizing a PCIe 4.0 interface with 64GB/s data transfer capability. 
\setlength\tabcolsep{2.5pt} 
\begin{table}[t]
    \centering
    \caption{Structural details of large-scale MoE-based models}
    \label{tab:moe_models}
        \begin{tabular}{l|c|c|c|c|c}
        \hline
            \multicolumn{1}{c|}{\textbf{Model}} & \textbf{Blocks} & \textbf{Experts} & \textbf{Top-k} & \multicolumn{1}{c|}{\textbf{Experts Params.}} & \multicolumn{1}{c}{\textbf{Total Params.}}\\
            \hline
            Mixtral 8x7B & 32 & 8 & 2 & 45.1B & 46.6B\\
            \hline
            Phi-3.5 MoE & 32 & 16 & 2 & 40.3B & 41.7B\\
        \hline
        \end{tabular}
\end{table} 

\paragraph{Implementation}
We build our inference engine, DAOP, using the Hugging Face Transformers library~\cite{wolf2020transformers}, a state-of-the-art platform for NLP supporting PyTorch, TensorFlow, and JAX.
Our experiments simulate real-time inference scenarios by setting the batch size to one. 
We integrate DAOP into the evaluation framework~\cite{eval-harness} to assess model accuracy on downstream tasks discussed previously.
In this study, \textbf{Expert Cache Ratio (ECR)} is defined as the proportion of available expert slots on the GPU relative to the total number of experts in the model. We adjust this ratio to assess inference speed and accuracy variations, demonstrating DAOP’s scalability across different cache sizes. 
Using the ShareGPT dataset~\cite{ShareGPT}, we calibrate dominant experts for the initial expert cache setup and quantify end-to-end inference performance by the number of tokens generated per second. This dataset is solely for initializing experts and differs from downstream tasks used for accuracy evaluation.
For power efficiency, we measure the average power consumption of the entire platform, encompassing both CPU and GPU, using a power and energy monitoring device.

\paragraph{Baselines}
We evaluate DAOP's performance against several baselines: MoE-OnDemand, DeepSpeed-MII~\cite{DeepSpeed_MII}, Mixtral-Offloading~\cite{eliseev2023fast}, and Fiddler~\cite{kamahori2024fiddler}.
DeepSpeed-MII is a memory-efficient Python library optimized for LLM inference. 
This library incorporates advanced features such as blocked KV caching and high-performance CUDA kernels to accelerate text generation.
Mixtral-Offloading and Fiddler discussed in Section~\ref{sec:related_work}, represent additional comparative frameworks. 
Pre-gated MoE~\cite{hwang2024pre} is only effective for MoE models with smaller-scale experts and requires substantial fine-tuning; hence, we use MoE-OnDemand as a more practical baseline for direct comparison. 
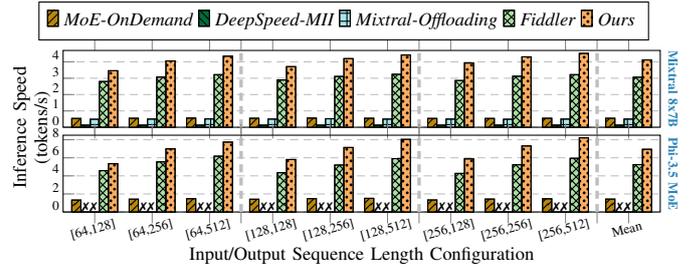
\begin{figure}[t]
    \centering
        \begin{tikzpicture}
    \centering
    \begin{groupplot}[
        group style={
            group name=my plots,
            group size=1 by 2,
            xlabels at=edge bottom,
            xticklabels at=edge bottom,
            vertical sep=3pt,
            horizontal sep=0pt,},
        xtick=data,
        x tick label style={font=\tiny,rotate=15,text height=0pt,text width=1cm,align=center,inner sep=0pt,outer sep=0pt},
        xticklabels from table={data/throughput_comp_mixtral.dat}{Config},
        xlabel style={font=\scriptsize, align=center, text height=0pt,inner sep=0pt,outer sep=-4pt},
        ytick pos=left,
        y tick label style={font=\tiny,rotate=0,text height=0pt,inner sep=1.5pt,outer sep=0pt},
        ylabel style={font=\scriptsize, align=center, text height=0pt,inner sep=0pt,outer sep=0pt},
        xlabel={Input/Output Sequence Length Configuration},
        ]

        \nextgroupplot[width=1.05\columnwidth, height=2.6cm,
                        ylabel={Mixtral 8x7B},
                        ylabel style={font=\tiny\bfseries, color=rendering, rotate=180, at={(1.016,0.5)}},
                        bar width=0.16,
                        ybar=0pt,    
                        ymin=0,
                        ymax=4.7,
                        ymajorgrids=true,
                        grid style=densely dashed,
                        enlarge x limits={abs=0.55},
                        ytick={0,1,2,3,4,5},
                        yticklabels={0,1,2,3,4,5},
                        extra x ticks={2.55,5.65,8.75},
                        extra tick style={grid=major,major grid style={densely dashed, line width=1.3pt}},
                        extra x tick labels={},
                        typeset ticklabels with strut,
                        legend columns=6,
                        legend style={at={(0.47,1.13,1.25)}, anchor=south, font=\scriptsize}
        ]
        
        \addplot[fill=darkgoldenrod,postaction={pattern=north east lines}]
        table[x expr={\coordindex + (ifthenelse(\coordindex >= 3, 0.1, 0) + ifthenelse(\coordindex >= 6, 0.1, 0) + ifthenelse(\coordindex >= 9, 0.1, 0))}, y = MoE-OnDemand,]{data/throughput_comp_mixtral.dat};
        \addlegendentry{\em MoE-OnDemand}
        
        \addplot[fill=cadmiumgreen,postaction={pattern=north west lines}]
        table[x expr={\coordindex + (ifthenelse(\coordindex >= 3, 0.1, 0) + ifthenelse(\coordindex >= 6, 0.1, 0) + ifthenelse(\coordindex >= 9, 0.1, 0))}, y = DeepSpeed-MII,]{data/throughput_comp_mixtral.dat};
        \addlegendentry{\em DeepSpeed-MII} 

        \addplot[fill=blizzardblue,postaction={pattern=grid}]
        table[x expr={\coordindex + (ifthenelse(\coordindex >= 3, 0.1, 0) + ifthenelse(\coordindex >= 6, 0.1, 0) + ifthenelse(\coordindex >= 9, 0.1, 0))}, y = Mixtral-Offloading,]{data/throughput_comp_mixtral.dat};
        \addlegendentry{\em Mixtral-Offloading}

        \addplot[fill=timedOutPointsColor,postaction={pattern=crosshatch}]
        table[x expr={\coordindex + (ifthenelse(\coordindex >= 3, 0.1, 0) + ifthenelse(\coordindex >= 6, 0.1, 0) + ifthenelse(\coordindex >= 9, 0.1, 0))}, y = Fiddler,]{data/throughput_comp_mixtral.dat};
        \addlegendentry{\em Fiddler}

        \addplot[fill=unsatPointsColor,postaction={pattern=crosshatch dots}]
        table[x expr={\coordindex + (ifthenelse(\coordindex >= 3, 0.1, 0) + ifthenelse(\coordindex >= 6, 0.1, 0) + ifthenelse(\coordindex >= 9, 0.1, 0))}, y = Ours,]{data/throughput_comp_mixtral.dat};
        \addlegendentry{\em Ours}

        \nextgroupplot[width=1.05\columnwidth, height=2.6cm,
                        ylabel={Phi-3.5 MoE},
                        ylabel style={font=\tiny\bfseries, color=rendering, rotate=180, at={(1.016,0.5)}},
                        bar width=0.16,
                        ybar=0pt,    
                        ymin=0,
                        ymax=8.5,
                        ymajorgrids=true,
                        grid style=densely dashed,
                        enlarge x limits={abs=0.55},
                        ytick={0,2,4,6,8},
                        yticklabels={0,2,4,6,8},
                        extra x ticks={2.55,5.65,8.75},
                        extra tick style={grid=major,major grid style={densely dashed, line width=1.3pt}},
                        extra x tick labels={},
                        typeset ticklabels with strut,
        ]
        
        \addplot[fill=darkgoldenrod,postaction={pattern=north east lines}]
        table[x expr={\coordindex + (ifthenelse(\coordindex >= 3, 0.1, 0) + ifthenelse(\coordindex >= 6, 0.1, 0) + ifthenelse(\coordindex >= 9, 0.1, 0))}, y = MoE-OnDemand,]{data/throughput_comp_phimoe.dat};
        
        \addplot[fill=cadmiumgreen,postaction={pattern=north west lines},nodes near coords={\tiny\xmark},nodes near coords style = {yshift=-0.8ex, xshift=0ex},]
        table[x expr={\coordindex + (ifthenelse(\coordindex >= 3, 0.1, 0) + ifthenelse(\coordindex >= 6, 0.1, 0) + ifthenelse(\coordindex >= 9, 0.1, 0))}, y = DeepSpeed-MII,]{data/throughput_comp_phimoe.dat};

        \addplot[fill=blizzardblue,postaction={pattern=grid},nodes near coords={\tiny\xmark},nodes near coords style = {yshift=-0.8ex, xshift=-0.1ex},]
        table[x expr={\coordindex + (ifthenelse(\coordindex >= 3, 0.1, 0) + ifthenelse(\coordindex >= 6, 0.1, 0) + ifthenelse(\coordindex >= 9, 0.1, 0))}, y = Mixtral-Offloading,]{data/throughput_comp_phimoe.dat};

        \addplot[fill=timedOutPointsColor,postaction={pattern=crosshatch}]
        table[x expr={\coordindex + (ifthenelse(\coordindex >= 3, 0.1, 0) + ifthenelse(\coordindex >= 6, 0.1, 0) + ifthenelse(\coordindex >= 9, 0.1, 0))}, y = Fiddler,]{data/throughput_comp_phimoe.dat};

        \addplot[fill=unsatPointsColor,postaction={pattern=crosshatch dots}]
        table[x expr={\coordindex + (ifthenelse(\coordindex >= 3, 0.1, 0) + ifthenelse(\coordindex >= 6, 0.1, 0) + ifthenelse(\coordindex >= 9, 0.1, 0))}, y = Ours,]{data/throughput_comp_phimoe.dat};
        
    \end{groupplot}   
    
    \draw (0,0) node[rotate=90, anchor=south] {\scriptsize \makecell{Inference Speed\\(tokens/s)}};
\end{tikzpicture}
        \caption{Inference speed comparison with full GPU memory utilization.}
        \label{fig:throughput_comp}
\end{figure} 

\subsection{Speedup}
Fig.~\ref{fig:throughput_comp} compares the inference performance of DAOP against baselines under various input/output sequence length configurations.
All methods aim to maximize GPU memory utilization, with an ECR of 46.9\% for MoE-OnDemand, Fiddler, and DAOP.
Due to significant migration overhead, baseline algorithms such as MoE-OnDemand, DeepSpeed-MII, and Mixtral-Offloading achieve less than one token per second for Mixtral 8x7B. 
Conversely, Fiddler and DAOP, which incur reduced expert migration overhead, demonstrate significant performance improvements that improve with increasing output token length. 
Notably, DAOP outperforms Fiddler by 40.4\%, achieving a token generation rate of 4.52 per second for Mixtral 8x7B and 8.21 per second for Phi-3.5 MoE with an input/output configuration of [256,512]. 
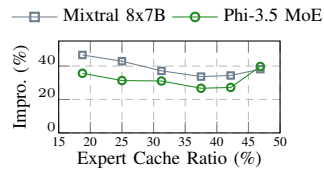
\begin{wrapfigure}{lo}{0.25\textwidth}
  \centering
    \begin{tikzpicture}
    \centering
    \pgfplotstableread[col sep=space,header=true]{
        Ratio Impro
        18.75	46.66666667
        25	42.92035398
        31.25	37.14285714
        37.5	33.7037037
        42.1875	34.375
        46.875	38.14102564
    }\mixtral
    \pgfplotstableread[col sep=space,header=true]{
        Ratio Impro
        18.75	35.69405099
        25	31.33159269
        31.25	31.06796117
        37.5	26.72605791
        42.1875	27.21649485
        46.875	39.77055449
    }\phimoe
    \begin{groupplot}[
        group style={
            group name=my plots,
            group size=1 by 1,
            xlabels at=edge bottom,
            xticklabels at=edge bottom,
            vertical sep=0pt,
            horizontal sep=0pt,},
        xtick=data,
        x tick label style={font=\tiny,rotate=0,text height=1pt,text width=1.5cm,align=center,inner sep=1pt,outer sep=3pt},
        xlabel style={font=\scriptsize, align=center, text height=2pt,inner sep=1pt,outer sep=1pt},
        ytick pos=left,
        y tick label style={font=\tiny,rotate=0,text height=0pt,inner sep=1pt,outer sep=0pt},
        ylabel style={font=\scriptsize, align=center, text height=0pt,inner sep=2pt,outer sep=0pt},]

        \nextgroupplot[width=0.5\columnwidth, height=2.8cm,
                        xlabel={Expert Cache Ratio (\%)},
                        ylabel={Impro. (\%)},
                        ymin=0,
                        ymax=55,
                        xmin=15,
                        xmax=50,
                        ymajorgrids=true,
                        xtick={15,20,25,30,35,40,45,50},
                        xticklabels={15,20,25,30,35,40,45,50},
                        grid style=densely dashed,
                        typeset ticklabels with strut,
                        legend columns=2,
                        legend style={at={(0.5,1.05,1.25)}, anchor=south, font=\scriptsize, draw=none,
                        legend image post style={mark size=1.6pt}},grid]

        \addplot+[mark=square, slategray, mark size=1.2pt, mark options={line width=0.8pt, solid, fill=slategray},]table[x index=0,y index=1,]\mixtral;
        \addlegendentry{Mixtral 8x7B}

        \addplot+[mark=o, deeperGreen, mark size=1.4pt, mark options={line width=0.8pt, solid, fill=deeperGreen},]table[x index=0,y index=1,]\phimoe;
        \addlegendentry{Phi-3.5 MoE}


        
            
    \end{groupplot}           

\end{tikzpicture}
    \caption{Inference speed improvement of DAOP over Fiddler for input/output length 256.}
    \label{fig:varied_ratio_comp}
\end{wrapfigure}
Fig.~\ref{fig:varied_ratio_comp} compares the performance of DAOP and Fiddler across various expert cache sizes using an input/output token length of 256.
The results show that DAOP consistently achieves an average performance improvement of 35.4\% over Fiddler. Even with only 25\% of experts cached on the GPU, DAOP achieves a speed of 3.23 tokens/s for the Mixtral 8x7B model and 5.03 tokens/s for the Phi-3.5 MoE model (see Fig.~\ref{fig:throughput_comp} for results at 46.9\% ECR).

\subsection{Energy Efficiency}
Table~\ref{tab:power_efficiency} compares the energy efficiency of DAOP with other baselines, showing that DAOP is the most energy-efficient method evaluated.
MoE-OnDemand, DeepSpeed-MII, and Mixtral-Offloading rely solely on the faster GPU for acceleration. However, frequent data transfers from CPU memory cause significant energy consumption, which is particularly severe for DeepSpeed-MII due to its lack of an efficient expert offloading mechanism. Conversely, Fiddler and DAOP leverage CPU and GPU resources for expert execution, reducing excessive expert loading and significantly outperforming these methods. Additionally, DAOP employs selective expert pre-calculation, favoring GPU execution for next-best experts, increasing GPU utilization, and achieving an average 1.50$\times$ improvement over Fiddler.
\setlength\tabcolsep{2pt} 
\begin{table}[t]
    \centering
    \caption{Energy efficiency (tokens/{\scriptsize k}J) comparison$^{\mathrm{a}}$}
    \label{tab:power_efficiency}
        \begin{tabular}{l|c|c|c|c|c}
        \hline
            \multicolumn{1}{c|}{\textbf{Model}} & \textbf{\makecell{MoE-\\OnDemand}} & \textbf{\makecell{DeepSpeed-\\MII}} & \textbf{\makecell{Mixtral-\\Offloading}} & \textbf{Fiddler} & \textbf{Ours}\\
            
            \hline
            Mixtral 8x7B & 2.63 & 0.59 & 2.13 & 10.06 & 14.37\\
            \hline
            Phi-3.5 MoE & 6.94 & -- & -- & 17.15 & 27.07\\
        \hline
        \multicolumn{6}{l}{$^{\mathrm{a}}$For input/output length 256 with full GPU memory utilization.}
        \end{tabular}
\end{table}

\subsection{Accuracy Results}
Table~\ref{tab:acc_compare_prefill} presents the model accuracy for various tasks using DAOP, focusing on input sequences during the prefill phase, specifically the first output token generated rather than the entire output sequence~\cite{eval-harness}.
The results confirm that our approximate optimizations during decoding do not affect task performance.
Additionally, Table~\ref{tab:acc_compare} explores the impact of DAOP on model accuracy throughout the entire inference phase with varying ECRs. 
The results show minimal impact on accuracy while maintaining efficiency, even with only 37.5\% of experts cached on the GPU. The only exception is  GSM8K dataset, which we discuss in the next section.
\setlength\tabcolsep{2pt} 
\begin{table}[t]
    \centering
    \caption{Impact of DAOP on model accuracy (\%) across downstream tasks dependent on the prefill stage}
    \label{tab:acc_compare_prefill}
        \begin{tabular}{l|c|c|c|c|c|c|c}
        \hline
            \multicolumn{1}{c|}{\textbf{Model}} & \textbf{Method} & \textbf{HellaS}& \textbf{Arc-e}& \textbf{Arc-c} & \textbf{PIQA} & \textbf{WinoG} & \textbf{MMLU} \\
            \hline
            \multirow{2}{*}{Mixtral 8x7B} & Official & 66.96 & 83.10 & 63.74 & 83.60 & 81.69 & 70.60 \\
            \cline{2-8}
            & Ours$^{\mathrm{a}}$ & 66.80 & 84.39 & 63.82	& 82.59	& 81.77 & 70.47 \\
            \hline
            \hline
            \multirow{2}{*}{Phi-3.5 MoE}  & Official & 69.21 & 76.77 & 66.64 & 78.84 & 78.37 & 78.78 \\
            \cline{2-8}
            & Ours$^{\mathrm{a}}$ & 69.25 & 76.43 & 66.38	& 79.00 & 78.37 & 78.69 \\
        \hline
        \multicolumn{8}{l}{$^{\mathrm{a}}$Our algorithm utilizes an Expert Cache Ratio (ECR) of 25.0\%.}
        \end{tabular}
\end{table}

\setlength\tabcolsep{2pt} 
\begin{table}[t]
    \centering
    \caption{Impact of DAOP on model accuracy (\%) across downstream tasks dependent on entire inference stage}
    \label{tab:acc_compare}
    \resizebox{\linewidth}{!}{
        \begin{tabular}{l|c|r|c|c|c|c|c}
        \hline
            \multicolumn{1}{c|}{\multirow{2}{*}{\textbf{Model}}} & \multirow{2}{*}{\textbf{Method}} & \multicolumn{1}{c|}{\multirow{2}{*}{\textbf{\makecell{ECR}}}} & \textbf{TriviaQA} & \textbf{BBH} & \multicolumn{2}{c|}{\textbf{TruthfulQA}} & \textbf{GSM8K} \\
            \cline{4-8}
            & & & \textbf{ExactMatch} & \textbf{ExactMatch} & \textbf{R1}$^{\mathrm{a}}$ & \textbf{R2}$^{\mathrm{a}}$ & \textbf{ExactMatch}\\
            \hline
            \multirow{5}{*}{Mixtral 8x7B}  & Official & 100.0\% & 71.59 & 49.36 & 45.04 & 38.43 & 58.91 \\
        
            \cline{2-8}
            & \multirow{4}{*}{Ours
            } & 62.5\%	& 70.98	& 47.63	& 46.02 & 41.74 & 51.48\\
            \cline{3-8}
            & & 50.0\% & 70.60 & 47.10 & 45.29 & 41.86 & 48.07\\
            \cline{3-8}
            & & 37.5\% & 70.13 & 47.14 & 48.10 & 42.72 & 41.77 \\
            \cline{3-8}
            & & 25.0\% & 69.08 & 46.61 & 48.47 & 44.31 & 33.51\\
            \hline
            \hline
            \multirow{5}{*}{Phi-3.5 MoE}  & Official & 100.0\% & -- &	56.11 &	56.30 &	39.66 & 86.88 \\
       
            \cline{2-8}
            & \multirow{4}{*}{Ours} & 62.5\% & --	& 56.54 & 54.59 & 39.53 & 82.79\\
            \cline{3-8}
            & & 50.0\% & -- & 56.43 & 53.12 & 39.53 & 81.27\\
            \cline{3-8}
            & & 37.5\% & -- & 56.52 & 56.92 & 39.41 & 81.27 \\
            \cline{3-8}
            & & 25.0\% & -- & 56.26 & 49.45 & 35.13 & 74.07\\
        \hline
        \multicolumn{8}{l}{\small$^{\mathrm{a}}$R1 and R2 denote the Rouge-1 and Rouge-2 scores, respectively.}
        \end{tabular}
    }
\end{table}

\section{Discussion}
\subsection{Platform and Model Applicability}
DAOP leverages parallel GPU-CPU execution by adaptively offloading non-critical experts to the CPU for selective pre-computation. 
Its enhanced inference speed and energy efficiency are based on the following assumptions: 1) GPU memory is limited for storing model weights. 2) The GPU is faster and more energy-efficient for model execution than the CPU. 3) CPU-GPU transfer latency exceeds the time required for expert execution on the CPU. 
Most commercial GPU devices~\cite{rtx4090, AMDMI300, h100} satisfy these assumptions, enabling DAOP to provide faster and more energy-efficient inference optimization.

\subsection{Limitation \& Future Work}
DAOP significantly accelerates inference with minimal accuracy trade-offs across most benchmarks, including commonsense reasoning, text generation, and world knowledge. 
However, for the GSM8K mathematics dataset, the diverse expert activations within a single sequence limit the capability of a small expert cache to capture activation patterns. We also measure the expert activation pattern variation during the decoding stage for each input sequence using a 15-token window. 
The results show that the cosine similarity for GSM8K decreases by 3.43\% compared to TriviaQA, validating our hypothesis.

\section{Conclusion}
Our proposed inference engine, DAOP, provides a viable solution for executing MoE models on resource-constrained devices by dynamically offloading non-critical experts to CPUs and selectively pre-calculating based on predictions. This strategy optimizes resource use, mitigating GPU memory constraints without compromising accuracy. In extensive tests with widely used Mixtral and Phi MoE models across diverse datasets, DAOP consistently outperforms traditional expert caching and prefetching techniques by 8.20$\times$ and offloading methods by 1.35$\times$. The code is publicly available at: \url{https://github.com/ecolab-nus/DAOP}.

\section*{Acknowledgment} 
We thank anonymous reviewers for their invaluable feedback that improved the work. This work is supported by the National Research Foundation, Singapore, under its Competitive Research Program Award NRF-CRP23-2019-0003.

\bibliographystyle{IEEEtran}
\bibliography{refs}

\end{document}